\documentstyle[12pt,world_sci]{article}
\def\overlay#1#2{\ifmmode%
\setbox0=\hbox{$#1$}%
\setbox1=\hbox to\wd0{\hss$#2$\hss}\else%
\setbox0=\hbox{#1}%
\setbox1=\hbox to\wd0{\hss#2\hss}\fi%
 #1\hskip-\wd0\box1 }

\begin{document}
\hfill\vbox{\hbox{\bf NUHEP-TH-94-19}\hbox{Aug 1994}}\par
\title{\bf PERTURBATIVE QCD FRAGMENTATION FUNCTIONS AS
A PHENOMENOLOGICAL MODEL FOR CHARM/BOTTOM  FRAGMENTATION
}
\author{KINGMAN CHEUNG\footnote{To appear in proceedings of DPF'94 meeting,
 Albuquerque, NM (August 1994)} \\
{\em  Department of Physics, Northwestern University,
Evantson, Illinois  60208, U.S.A.\\}
}
\vspace{0.3cm}

\maketitle
\setlength{\baselineskip}{2.6ex}

\begin{center}
\parbox{13.0cm}
{\begin{center} ABSTRACT \end{center}
{\small \hspace*{0.3cm}
The perturbative QCD fragmentation functions
can be applied phenomenologically as a  model
for charm and bottom quark fragmentation into heavy-light mesons.  The
predictions by this model on the observables $P_V$ and $\langle z \rangle$
for $D-D^*$ and $B-B^*$ systems are compared with experimental data.
}}
\end{center}

The dominant production mechanism for
mesons and baryons that contain a single heavy quark
is the fragmentation of the heavy quark,
in which light quark-antiquark pairs are created out of the vacuum by the color
force of the heavy quark and then the heavy quark captures the light quarks to
form mesons or baryons.  However, in this process, the creation of light
quark-antiquark pairs tells us that the nonperturbative effects are important
and so the fragmentation function, which describes the process, cannot be
calculated from the first principle or from perturbative QCD.

Recently, there are new developments in the fragmentation of heavy quarks into
heavy-heavy-quark bound states \cite{gluon,jpsi,bc}.  The
fragmentation process involves the creation of heavy
quark-antiquark pair, which implies that the natural scale of the process
should be of order of the mass of the heavy quark created, and so the process
should be calculable by perturbative QCD (PQCD).
Fragmentation functions for $c\to (c\bar c)$, $b\to (b\bar b)$, and $\bar b\to
(\bar b c)$ in different spin-orbital states have been calculated to leading
order in $\alpha_s$ using PQCD.
In addition to the application that these PQCD fragmentation functions can
predict the production rates of heavy-heavy-quark bound states without model
dependence, they can also be applied as a phenomenological model to describe
the charm and bottom quark fragmentation into heavy-light mesons
\cite{bc-pol,eff}.

The derivation of the $\bar b\to B_c, B_c^*$ fragmentation functions was given
elsewhere \cite{bc,bc-pol}.  Here we quote the results:
\begin{eqnarray}
\label{dz1}
D_{\bar b\rightarrow B_c}(z,\mu_0) & = &
N \, \frac{rz(1-z)^2}{(1-(1-r)z)^6}
\left[ 6 - 18(1-2r)z + (21 -74r+68r^2) z^2 \right. \nonumber \\
 && \left. -2(1-r)(6-19r+18r^2)z^3  + 3(1-r)^2(1-2r+2r^2)z^4 \right]
\end{eqnarray}
for the $^1S_0$ state and
\begin{eqnarray}
\label{dz2}
D_{\bar b\rightarrow B_c^*}(z,\mu_0) & = &
3N\,  \frac{rz(1-z)^2}{(1-(1-r)z)^6}
\left[ 2 - 2(3-2r)z + 3(3 - 2r+ 4r^2) z^2  \right . \nonumber \\
&& \left.  -2(1-r)(4-r +2r^2)z^3  + (1-r)^2(3-2r+2r^2)z^4 \right]
\end{eqnarray}
for the $^3S_1$ state, where
$N=2\alpha_s(2m_c)^2 |R(0)|^2/(81 \pi m_c^3)$ and $r=m_c/(m_b+m_c)$.

The fragmentation functions in Eqs.~(\ref{dz1}) and (\ref{dz2}) can also be
regarded as  two-parameter functions with $N$ and $r$ as free parameters.  $N$
governs the overall normalization and $r=m_{\rm light}/m_{\rm meson}$ is the
mass ratio.   We vary the parameter $r$ to study the behavior of these
fragmentation functions in the limit $r=m_{\rm light}/m_{\rm meson} \to 0$.
We expect that in this limit these fragmentation
functions can describe, to certain extent,  the fragmentation of the heavy
quark into heavy-light mesons, {\it e.g.},
$c\to D,\,D^*$ and $\bar b\to B,\,B^*$.
Although in the fragmentation $c\to D,\,D^*$ and $\bar b\to B,\,B^*$ there
are probably large nonperturbative and relativistic effects that we have not
taken into account, the PQCD fragmentation functions with the free parameters
$N$ and $r$ can at least provide some insights to these systems while precise
nonperturbative fragmentation functions for charm and bottom are not available
yet.

\begin{figure}[t]
\vspace{3in}
\caption{\label{fig1}
(a) The ratio $P_V=D^*/(D+D^*)$ for the $D-D^*$ system predicted by the
PQCD fragmentation functions, and the comparison with the data;
(b) the average $\langle z \rangle^\mu$ for $c\to D^*$ and for $\bar b\to B^*$
fragmentation versus the scale $\mu$.  The experimental measurements from
LEP ($\mu=m_Z/2$) and CLEO/ARGUS ($\mu=5.3$ GeV) are also shown.
}
\end{figure}

The ratio $P_V$ for the $D$ and $D^*$ system is defined as
$P_V =\frac{D^*}{D+D^*}$,
which is a measure of the relative population of $D^*$ in the production of
$D$ and $D^*$ mesons.   Since fragmentation is the dominant production
mechanism, the production rates of $D$ and $D^*$ can be
replaced by the corresponding fragmentation probabilities as
\begin{equation}
P_V= \frac{P_{c\to D^*}}{P_{c\to D} + P_{c\to D^*} }\;,
\end{equation}
which is a function of $r$ only.
The probabilities $P_{c\to D}$ and $P_{c\to D^*}$ are  obtained by
integrating $D_{c\to D}(z)$ and $D_{c\to D^*}(z)$ over $z$.
The prediction by our fragmentation model is
shown in Fig.~\ref{fig1}(a).   At $r=0$, which is the heavy quark mass limit,
the ratio $P_V=0.75$ is exactly the value given by heavy quark spin symmetry.
The average experimental value $P_V=0.646\pm0.049$.
For the charm system we choose $m_c=1.5$ GeV and the light constituent quark
mass $m_{u,d}$ inside the $D$ and $D^*$ mesons to be 0.3 GeV, therefore
$r=0.167$.  The experimental data point $(r=0.167, P_V=0.646\pm0.049)$ is
plotted in Fig.~\ref{fig1}(a).  We found good agreement between the
prediction of our fragmentation model and the data.
Physically, $P_V<0.75$ means that the production rate of
$D^*$ is less than it should be as given by heavy quark spin symmetry.
For the $B-B^*$ system, with the same value for $m_{u,d}$ and
$m_b=4.9$ GeV, we have $r=0.058$.  The value of $P_V$ at $r=0.058$ is about
0.68, which is closer to the heavy-quark-symmetry prediction of 0.75 than the
$D-D^*$ system.

The $\langle z \rangle$ is the average longitudinal momentum fraction that is
transferred from the heavy quark to the meson.  In terms of fragmentation
functions, $\langle z \rangle^{\mu}$ at a scale $\mu$ for $c\to D^*$
fragmentation is given by
\begin{equation}
\label{z}
\langle z \rangle^{\mu}_{c\to D^*} = \frac{\int dz\, zD_{c\to D^*}(z,\mu)}
{\int dz \, D_{c\to D^*}(z,\mu)}\;.
\end{equation}
The scaling behavior of $\langle z \rangle^{\mu}$ is given by
$\langle z \rangle^{\mu} = \langle z \rangle^{\mu_0} [
\alpha_s(\mu)/ \alpha_s(\mu_0) ]^{-2\gamma/b}$,
where $\gamma=-4C_F/3$, $C_F=4/3$, $b=(11N_c - 2 n_f)/3$, $N_c=3$, $n_f$ is
the number of active flavors at the scale $\mu$, and $\langle z
\rangle^{\mu_0}$ is the value determined at the initial scale $\mu_0$.
Taking the inputs: $m_c=1.5$ GeV, $m_{u,d}=0.3$ GeV, and
$\mu_0=m_c+2m_{u,d}=2.1$ GeV, we have $r=0.167$,
$\langle z \rangle^{\mu_0}_{c\to D^*}=0.77$, and
$\langle z \rangle^{\mu=m_Z/2}_{c\to D^*}=0.50$.
For $\bar b \to B^*$ fragmentation, we choose $m_b=4.9$ GeV, $m_{u,d}=0.3$
GeV, and $\mu_0=m_b+2m_{u,d}=5.5$ GeV, we have $r=0.058$,
$\langle z \rangle^{\mu_0}_{\bar b\to B^*}=0.87$, and
$\langle z \rangle^{\mu=m_Z/2}_{\bar b\to B^*}=0.696$.
The curves for
$\langle z \rangle^{\mu}_{c\to D^*}$ and
$\langle z \rangle^{\mu}_{\bar b\to B^*}$ versus $\mu$ are shown in
Fig.~\ref{fig1}(b).

Experimental data on $\langle z \rangle$
are available from the LEP detectors and from CLEO and ARGUS detectors.
The LEP average
$\langle x_E \rangle_{c\to D^*} = 0.504 \pm 0.0133$.
For the bottom quark,  only the inclusive hadron production has been measured
and $\langle x_E \rangle_{b\to H_b}=0.694\pm 0.0166$.
The combined CLEO and ARGUS  data is
$\langle x_E \rangle_{c\to D^*}=0.648\pm0.043$.  The scale of the measurements
is taken to be one half of the center-of-mass energy of the machines.
These data are shown in
Fig.~\ref{fig1}(b).  Excellent agreement is demonstrated.

Here, we have demonstrated the applications of the perturbative
 QCD fragmentation functions as a phenomenological model to describe the
fragmentation of heavy quarks into heavy-light mesons and the agreement
between the predictions by this model and the experimental data.

I am indebt to my collaborators: Tzu Chiang Yuan and Eric Braaten.
This work was supported by the DOE Grants DE-FG02-91-ER40684.

\bibliographystyle{unsrt}

\end{document}